
\input phyzzx.tex
 \PHYSREV


%
\def\M{{\cal M}}

\def\kp{k_{\perp} }

\def\order#1{{\cal O}(#1)}
\def\Jpsi{{J/\psi}}

\def\epem{{e^+ e^-}}
\def\half{{\textstyle {1\over 2}}}
\def\GeV{{\rm GeV}}

\def\PL #1 #2 #3 {Phys.~Lett.~{\bf#1}\ (#2) #3}
\def\NP #1 #2 #3 {Nucl.~Phys.~{\bf#1}\ (#2) #3}
\def\PR #1 #2 #3 {Phys.~Rev.~{\bf#1}\ (#2) #3}
\def\PP #1 #2 #3 {Phys.~Rep.~{\bf#1}\ (#2) #3}
\def\PRL #1 #2 #3 {Phys.~Rev.~Lett.~{\bf#1}\ (#2) #3}
\def\ZP #1 #2 #3 {Z..~Phys.~{\bf#1}\ (#2) #3}


\overfullrule 0pt


\def\etal{{\it et. al.}}

 \doublespace
 \Pubnum{HU-TFT-92-37}
 \date{October 1992}

 \titlepage
\vfill
 \title{%
NEW QCD EFFECTS AT LARGE $x$%
 \foot{Lectures presented at the XXXII Cracow School of Theoretical Physics,
Zakopane 1992.}}
 \author{Paul Hoyer}
\address{Department of Physics \break 
University of Helsinki, Helsinki, Finland}
\vfill

\abstract

In heavy quark production at large Feynman $x$ there are two hardness
scales, one given by the heavy quark pair mass $\M^2$\ and the other by
$\Lambda_{QCD}^2/(1-x)$. When these two scales are comparable, the
twist expansion of Perturbative QCD breaks down. We discuss the dynamics in
this new QCD limit, where $\mu^2=\M^2(1-x)$ is held fixed as $\M^2\to
\infty$. New diagrams are found to contribute, which can enhance the cross
section above that expected for leading twist. The heavy quarks are produced
by a peripheral scattering on the target of hardness $\mu^2$. This leads, in
particular, to a nuclear target dependence of $A^{2/3}$ at small $\mu$.
Qualitatively, the dynamics in the new limit agrees with earlier
phenomenological models of ``intrinsic'' heavy quark production.

\vfill
\endpage

\REF\fac{
J. C. Collins, D. E. Soper and G. Sterman, published in {\it Perturbative
QCD}, edited by A.H. Mueller, World Scientific, (1989);
G. Bodwin, Phys. Rev. {\bf D31} (1985) 2616 and {\bf D34}  (1986E) 3932;
J. Qiu and G. Sterman, Nucl. Phys. {\bf B353} (1991) 105, {\it ibid.}, 137.}

\chapter{Introduction}

The application of Perturbative QCD (PQCD) to hard processes involving
hadrons and leptons has been very successful.  The predictive power of PQCD
is based on the factorization theorem\refmark\fac, according to which an
observable cross-section $\sigma$ can be factorized into a product of
universal structure and fragmentation functions times the hard constituent
cross-section ${\hat\sigma}$. Generically,
$$\sigma = F_a F_b\,{\hat\sigma}(ab \to c+X)D_c \eqn\fact$$
where $F_a, F_b$ are the single
parton structure functions of the incoming hadrons (``probabilities for
finding the quarks/gluons $a,b$ in the projectile and target'') and $D_c$
describes the hadronization of the  produced quark(s)/gluon(s) $c$ into the
observed final state (\eg, jets or hadrons at large $p_\bot$). The
structure and fragmentation functions  $F, D$ are not calculable in PQCD,
but are predicted to be {\it universal,}  \ie, independent of the specific
hard collision ${\hat\sigma}$.  The ``higher twist'' corrections to the
``leading twist'' PQCD prediction (1) are suppressed by powers of the large
momentum scale $Q^2$, which characterizes the hard subprocess $ab\to c+X$.

\REF\bhmt{
S. J. Brodsky, P. Hoyer, A. H. Mueller, W.-K.
Tang, Nucl. Phys. {\bf B369} (1992) 519.}

\REF\exc{
S. J. Brodsky and G. P. Lepage, published in {\it Perturbative
QCD}, edited by A.H. Mueller, World Scientific, (1989).}

Here I would like to discuss modifications to Eq. \fact\ that are
expected, and observed, in the case that one of the partons
$a, b$ or $c$ carries a  large fraction $x$ of the available longitudinal
momentum. In the $x\to 1$ limit there is a new hard scale
$\Lambda_{QCD}^2/(1-x)$, and the corrections to eq. \fact\ are of order
$\Lambda_{QCD}^2/Q^2(1-x)$.  In the combined limit
$$\left.\eqalign{Q^2 &\to \infty\cr
x &\to 1\cr}\right\}
\eqalign{&{\hskip 1cm} {\rm\ with\ } \mu^2=Q^2(1-x) {\rm\ fixed}\cr}
\eqn\xlimit$$
the twist expansion in fact breaks down\refmark\bhmt.  The higher twist
contributions are of $\order{1/\mu^2}$, and hence not suppressed by an
asymptotically large variable in this new QCD limit.   Since large momentum
transfers are involved, PQCD can still be used to analyze the process.  As
I shall discuss below, in this limit the cross-section cannot be expressed
in terms of single-parton structure functions; it involves multiparton
distributions.  This is similar to the case of exclusive $(x=1)$ hard
scattering, which depends on the longitudinal  momentum distributions of all
(valence) quarks, constrained to be at  small transverse distances from each
other\refmark\exc.

\chapter{Why the $x\to 1$ Limit is Hard}

Consider the wave function $\phi(yp,{\vec n}_\bot)$ of the hadron $h$ in
Fig.~1, in a frame where $h$ has a large longitudinal momentum $p$. We
shall assume that $\phi$ describes the soft, non-perturbative part of the
quark distributions, and is suppressed in the limits $y\to 0,1$ and also for
$n_\bot \to \infty$.  The perturbative tail of the full wave
function\footnote*{The following simplified discussion ignores multiple
gluon exchanges, which will bring logarithmic corrections, in analogy to
the treatment in Ref.\refmark\exc.} when the fractional momentum of one
constituent $x\to 0,1$ or when its transverse momentum $p_\bot \to \infty$
is then generated by gluon exchange as shown in Fig.~1.

It is in fact straightforward to see\refmark\bhmt\
that Fock states where one constituent carries a large momentum fraction
$x\simeq 1$ must have a short life-time, and hence are calculable in PQCD.
The (kinetic) energy difference between the final Fock state of Fig.~1
and the hadron is $$\eqalign{\Delta E_{q\bar q}&\equiv E_h - E_{q\bar q} =
\sqrt{p^2+m_h^2} - \sum_i \sqrt{(x_ip)^2+p_\bot^2+m_q^2}\cr
 &\simeq {1\over {2p}} \left[m_h^2 - {m_q^2+p_\bot^2\over x(1-x)}\right]
  \propto {1\over 1-x}} \eqn\delqq$$
 where we assumed $p>>p_\perp/(1-x)$.  By the uncertainty principle the
``life-time''
 $\tau_{q\bar q}$ of the Fock state is then proportional to $1-x$:
 $$\tau_{q\bar q} \simeq {1\over \Delta E} \simeq {2px \over m_q^2+p_\bot^2}
 (1-x) \eqn\tauqq$$

 Furthermore, since the life-time of the $x\to 1$ Fock state is brief, we
would expect that the transverse distance $r_\bot$ in Fig.~1, between the
quarks
 before/during the gluon exchange must be similarly short, so that the
 duration of the exchange is no longer than the life-time $\tau_{q\bar q}$.
 This is readily verified
 because the gluon exchange amplitude depends on the constituent
transverse
 momentum ${\vec n}_\bot$ only through the energy difference associated
 with the dashed line in Fig.~1,
 $$2p\Delta E \simeq {m_q^2+n_\bot^2\over 1-y} - {m_q^2+p_\bot^2\over 1-x} -
 {({\vec n_\bot-\vec p}_\bot)^2\over x-y} \eqn\dele$$
 For $x\to 1$, $\Delta E$ is independent of $n_\bot$ for
 $$n_\bot^2 \lsim {\cal O}(p_\bot^2/(1-x)) \eqn\nper$$
 Since the non-perturbative wave function $\phi$ of the hadron will
 cut off the integration over ${\vec n}_\bot$ before the limit \nper\ is
 reached, the $n_\bot$-integration can be factorized,
 $$\int d{\vec n}_\bot \phi(yp, {\vec n}_\bot) = \phi(yp, r_\bot=0)
 \eqn\nint$$
 showing that the only hadronic Fock states which can generate the $x\to 1$
 perturbative tail are those with a short transverse distance $r_\bot^2
 \propto 1-x$ between the quarks.  Conversely, according to Eq. \nper\ the
 $x\to 1$ Fock states involve large transverse momenta
 $n_\bot^2 \propto 1/(1-x)$.

 It is important to note that the $x\to 1$
 Fock state, which is compact at its moment of creation,
 nevertheless quickly expands in the transverse direction.
 During the
 effective life-time \tauqq\ of the Fock state, the ``slow'' quark can
 move a transverse distance
 $$R_\bot \simeq v_\bot \tau_{q\bar q} = {p_\bot\over p(1-x)} {2px\over
 m_q^2+p_\bot^2} (1-x) \simeq {2p_\bot x\over m_q^2+p_\bot^2}\eqn\bperp$$
which for $p_\bot = {\cal O}(\Lambda_{QCD})$ is of the order of 1 fm.
 This observation will be important in the following.

\REF\spect{%
D. Sivers, S. J. Brodsky and R. Blankenbecler,
Phys. Reports {\bf 23C} (1976) 1.}

\REF\gnb{J. F. Gunion, P. Nason and R. Blankenbecler, {Phys.~Rev.~{\bf D29}\
(1984) 2491}.}

\REF\bs{S. J. Brodsky and I. Schmidt, Phys.~Lett. {\bf B234} (1990) 144.}

 Because of the hard scale \nper, PQCD can be used to calculate the behavior
 of hadronic structure functions in the $x\to 1$ limit.  The result is given
 by the ``spectator counting rules''\refmark{\spect,\gnb,\bs}
 $${dF\over dx} \propto
(1-x)^{2n_s-1+2\vert\Delta\lambda\vert}\eqn\dcr$$
 where $n_s$ is the number of ``spectator'' partons (whose $x_s \to 0$)
 and $\Delta\lambda$ is the helicity flip between
 the initial hadron and the observed quark (or gluon) carrying the large
momentum fraction $x$.

\chapter{Breakdown of the Twist Expansion for $x\to 1$}

 Consider the well-known process of Deep Inelastic Scattering (DIS) of
leptons on hadrons.  The diagrams are classified as ``Leading Twist''
(Fig. 2a)
 or ``Higher Twist'' (HT, Fig. 2b) according to whether the spectator
 partons in the target are, or are not, connected to the active (hit)
 quark (or to partons radiated from this active quark).

 In the $Q^2 \to \infty$ (fixed $x$) limit, the HT diagrams are suppressed
 by $1/Q^2$.  This can be intuitively understood as follows.
 The hard $\ell q\to \ell^\prime q$ subprocess has a duration
 $\tau \sim 1/Q$.  Any interaction with spectators as in Fig. 2b can
 affect the hard scattering probability only if it occurs within this short
 time interval $\tau$ (later interactions will only modify the momentum
 distribution of the struck quark in the final state).  But an interaction
 within a time-scale $\tau$ is possible only if there are spectators
 within a transverse distance $r_\bot \sim \tau \sim 1/Q$.  The probability
 for this is proportional to the transverse area $\pi\,r_\bot^2 \sim 1/Q^2$,
 which explains the suppression of the higher twist contributions.

 The above argument for the suppression of HT terms breaks down in the
 high $x$ limit.  As we argued earlier, the $x \to 1$ Fock states in Fig.
 1 are produced from compact hadron configurations, with a typical
 distance $r_\bot^2 \sim (1-x)/\mu^2$ between the valence quarks, where
 $\mu \sim 1\ {\rm{fm}}^{-1}$.  Hence in DIS with $Q^2 \sim \mu^2/(1-x)$,
 \ie, in the limit \xlimit, the scale of the hard photon interaction is
 commensurate with the size of the valence Fock state, and the scattering
 is coherent over several quarks.  This means that the DIS cross-section
 cannot be expressed in terms of single-parton structure functions in the
 limit \xlimit, and the usual factorization \fact\ fails.

 \REF\vm{M. Virchaux and A. Milsztajn, Phys.~Lett.~{\bf B274}\ (1992) 221.}

 \REF\sol{M. Soldate, {Nucl.~Phys.~{\bf B223}\ (1983) 61.}}
 \REF\naten{S. Falciano, \etal, Z.~Phys. {\bf C31} (1986) 513; {\it ibid.,}
 {\bf C37} (1988) 545.}
 \REF\cp{J. S. Conway, \etal, Phys.~Rev. {\bf D39} (1989) 92.}

\REF\ht{E. L. Berger and S. J. Brodsky, {Phys.~Rev.~Lett.~{\bf 42}\ (1979)
940}; S. S. Agaev, Phys.~Lett. {\bf B283} (1992) 125.}

 The magnitude of the higher twist terms has been studied experimentally
 in DIS as a function of $x$. The results show\refmark\vm\ that the HT
corrections are
 important for $x \gsim 0.5$, and have an $x$-dependence which is consistent
 with $(1-x)^{-1}$, as suggested by the above qualitative arguments and
explicit calculations\refmark{\fac,\gnb,\sol}. Similarly, in high mass lepton
pair production there is experimental evidence\refmark{\naten,\cp} for
corrections to the Drell-Yan process, which are in qualitative agreement
with the expectations for higher twist effects\refmark{\ht}.

\chapter{Dynamics in the new QCD Limit}

 We have studied\refmark\bhmt\ the production of quark pairs with large
invariant mass
 ${\cal M}$ in the high $x$ limit corresponding to \xlimit, \ie,
 for
 $$\left.\eqalign{\M^2 &\to \infty\cr
x &\to 1\cr}\right\}
\eqalign{&{\hskip 1cm} {\rm\ with\ } \mu^2=\M^2(1-x) {\rm\ fixed}\cr}
\eqn\mlimit$$
 where $x$ is the momentum fraction of the $q\bar q$ system.  An explicit
 calculation of all relevant diagrams was done for scalar QED.
 Here I would like to discuss the qualitative conclusions, which apply
 equally to QCD.

 Consider first heavy quark production in the standard QCD limit,
 $${\cal M}^2 \to \infty \qquad {\rm at\ fixed\ } x.\eqn\qlimit$$
 The usual lowest order diagram describing the fusion process $GG\to c{\bar
c}$ is shown in Fig. 3a. The virtualities of both gluons range up to
 ${\cal O}({\cal M}^2)$ -- their momentum distributions are given by the
projectile and target
 gluon structure functions evaluated at a scale $Q^2 = {\cal M}^2$.
 Similarly in Fig. 3b, the higher order (in $\alpha_s$) process
 ${\bar q}G\to {\bar q}Q{\bar Q}$ is given by the antiquark and gluon
 structure functions of the projectile and target, respectively, evaluated
 at a scale ${\cal M}^2$.  Both diagrams in Fig. 3 are leading twist --
 they involve only one of the partons (a gluon and an antiquark,
 respectively) in the projectile, and one gluon in the target.

 The dynamics of the new limit \mlimit\ differs in several respects from
 the above.  There are two types of leading order diagrams, the ``extrinsic''
 and ``intrinsic'' ones shown in Figs. 4a and 4b, respectively. In
extrinsic diagrams the heavy quark pair couples to only one parton in the
projectile, while in intrinsic diagrams it couples to several. Since
 the produced $Q{\bar Q}$ pair carries almost all of the momentum in the
final state $(x\to 1)$, the light valence quarks  $q, {\bar q}$ are
effectively stopped.  The light quarks then give a big  contribution to the
energy of the intermediate states, $2p\Delta E \sim p_\bot^2/(1-x)$ (cf.
Eq. \dele).   The production cross section is dominated by values of the
light quark transverse momentum $p_\bot$ where this light quark contribution
to the energy difference is of the same order as that of the  heavy pair,
$2p\Delta E \sim {\cal M}^2$.  Hence (for $\mu\gsim \Lambda_{QCD}$)
$$p_\bot^2 \sim {\cal M}^2(1-x) = \mu^2 \eqn\pperp $$
Because of their small fraction $\sim (1-x)$ of the projectile momentum, the
stopped light quarks can move a considerable distance $R_\bot$ in the
transverse direction, even during the brief life-time of the virtual
$q{\bar q}Q{\bar Q}$ Fock state.  According to Eq. \bperp, $R_\bot \sim
1/p_\bot \sim 1/\mu$.  Now a Fock state of this transverse size can be
``resolved'' by a target gluon of transverse momentum $\ell_\bot$ (or
virtuality $\ell^2 \sim - \ell_\bot^2$) of order\footnote* {{\tenrm
Equivalently, we could say that an interaction of this hardness with the
target deflects the stopped light quarks sufficiently to break up the Fock
state and materialize the heavy quark pair.}} $\ell_\bot \sim 1/\mu$. Hence
the hardness of the scattering from the target does {\it not} increase with
the heavy quark pair mass \M.  To be able to describe the target scattering
using PQCD, we would have to choose $\mu >> \Lambda_{QCD}$  in the limit
\mlimit. In general, however, we must conclude that arbitrarily heavy quarks
can be, and are, produced at high $x$ by {\it soft} peripheral scattering.
Such soft scattering is surface-dominated for nuclear targets, \ie,
$\sigma \propto A^{2/3}$ as observed for $J/\psi$ production at large $x$
(see below).

Note that soft scattering is kinematically allowed only at sufficiently
high energies. The minimum longitudinal momentum transfer from a stationary
target required to put a heavy quark pair of mass $\M$ on its mass shell is
of order $\M^2/2p$, where $p$ is the laboratory momentum of the
projectile. For charm production the minimum momentum transfer is below
50~MeV already for $p > 100$ GeV.

The large transverse size $R_\bot$ of the light quark distribution also
explains why the scattering dominantly occurs off the light quarks, as shown
in Fig. 4.  The heavy $Q{\bar Q}$ pair has a small transverse size
$h_\bot \sim 1/{\cal M}$.  A target gluon can couple to, and
resolve\footnote*{If the virtual $Q\bar Q$ pair is in a color octet
configuration it will also interact coherently with soft gluons. In
this case the virtual pair behaves like pointlike gluon, and is not
materialized.}, the  $Q{\bar Q}$ pair only provided it has a commensurate
wavelength, \ie, $\ell_\bot \sim {\cal M}$ as indicated in Fig. 3a.  This is
much larger  than the $\ell_\bot \sim \mu$ required to resolve the light
quarks. Hence the gluon fusion diagram of Fig. 3a, which gives the leading
contribution in the usual, fixed $x$, QCD limit \qlimit, is actually
suppressed by $1/{\cal M}$ compared to the diagrams of Fig. 4 in the new
limit \mlimit.

The Fock state of the projectile hadron from
which the  heavy pair is produced must have a small transverse size
$r_\bot^2 \sim (1-x)/\mu^2 \sim 1/{\cal M}^2$ (\cf\ Fig. 4).  The
argument is the same as the one already given in Section 2, leading to the
estimate \nper\  of the valence quark transverse momentum before the virtual
creation of the heavy quark pair.

The extrinsic (Fig. 4a) and intrinsic (Fig. 4b) diagrams are of
the same order in $\alpha_s$ and have the same behavior
in the limit  \mlimit. However, we note some qualitative
distinctions between these two classes of diagrams:

(i) The intrinsic diagrams do not contribute significantly to lepton pair
production, since two photon exchanges would be required.  This may explain
the very different A-dependence of lepton pairs, as compared to the $J/\psi$
(see below).

\REF\bii{C. Biino {\it et al.}, Phys. Rev. Lett. {\bf 58} (1987) 2523.}

\REF\polar{%
D. L. Adams, \etal, Phys.~Lett.~{\bf B261} (1991) 201; {\it ibid.,}
{\bf B264} (1991) 462.}

(ii) The intrinsic diagrams give rise to a non-trivial phase in the leading
order amplitude. As shown in Fig. 5, in an intrinsic diagram the $Q\bar
Q$ production can proceed through two consecutive real processes. First an
on-shell $Q\bar Q$ pair with $x<1$ is formed, and later the pair is
accelerated to $x\to 1$ via an interaction with the second valence quark.
This dynamical phase can be of importance in polarization phenomena, which
have been observed\refmark{\bii,\polar} to be enhanced at large $x$.

(iii) The extrinsic diagrams dominate over the intrinsic ones both in the
exclusive ($\mu\ll \Lambda_{QCD}$ in Eq. \mlimit) and inclusive ($\mu\gg
\Lambda_{QCD}$) limits. In the extrinsic diagram of Fig. 4a, the
virtuality of the gluon exchange between the light valence quarks is set by
$1/r_\perp^2 \sim  \Lambda_{QCD}^2/(1-x)$, while that of the gluon
connecting to the heavy quark pair is set by $\M^2$. In the intrinsic
diagram of Fig. 4b, on the other hand, {\it both} gluons have
virtualities determined by the {\it larger} of these scales. In the scalar
QED model calculation of Ref. \bhmt, the intrinsic diagrams nevertheless
dominated at intermediate values of $\mu^2$. It would clearly be important to
determine the magnitude of the intrinsic contribution in QCD.

\chapter{The Production of Charmed Hadrons at Large $x$}

\REF\der{
M. Derrick, {\it Proceedings of the XXIV International
Conference on High Energy Physics},
(R. Kotthaus and J. H. K\"uhn, Eds., Springer 1989), p. 895.}

\REF\anj{J. C. Anjos, \etal, {Phys.~Rev.~Lett.~{\bf 62}\ (1989) 513.}}

\REF\chr{S. Bethke, {Z.~Phys.~{\bf C29}\ (1985) 175};
W. Bartel, \etal, {Z.~Phys.~{\bf C33}\ (1987) 339};
J. Chrin, {Z.~Phys.~{\bf C36}\ (1987) 163};
D. Decamp, \etal, {Phys.~Lett.~{\bf 244B}\ (1990) 551}.}

\REF\pet{C. Peterson, D. Schlatter, I. Schmitt, and P. Zerwas,
Phys. Rev. {\bf D27} (1983) 105.}

The data on charm production in $\epem$ annihilations\refmark\der\ and in
photoproduction\refmark\anj\ agrees well with PQCD at leading twist. The
charm fragmentation function $D(c\to D+X)$ can thus be determined from these
reactions using the factorization formula \fact. The $\epem$ data
show\refmark\chr\  that the
charmed hadron carries an average fraction $\VEV{z}\simeq 70$\% of the
charm quark momentum. The momentum distribution is often parametrized in
terms of the ``Peterson'' function\refmark\pet,
$$D_{H/c}(z) \propto {1\over {z(1 - 1/z -\epsilon_c/(1-z))^2}}\eqn\pete$$
where $\epsilon_c\simeq 0.06$. According to the QCD factorization theorem,
this fragmentation function should be the same in all hard processes,
regardless of how the charm quark is produced.

\REF\bia{S. F. Biagi, \etal, {Z.~Phys.~{\bf C28}\ (1985) 175}.}
\REF\cha{P. Chauvat, \etal, {Phys.~Lett.~{\bf 199B}\ (1987) 304}.}
\REF\cot{P. Coteus, \etal, {Phys.~Rev.~Lett.~{\bf 59}\ (1987) 1530}.}
\REF\shi{C. Shipbaugh, \etal, {Phys.~Rev.~Lett.~{\bf 60}\ (1988) 2117}.}
\REF\agu{M. Aguilar-Benitez, \etal, {Z.~Phys.~{\bf C40}\ (1988) 321}.}

\REF\app{J. A. Appel, FERMILAB-Pub-92/49, to appear in Ann. Rev. Nucl.
Part. Sci. {\bf 42} (1992);
G. A. Alves, \etal, FERMILAB-Pub-92/208-E (August, 1992) and Contribution
to ICHEP 92 (Dallas, Texas, August 1992).}
\REF\wa{M. Adamovich, \etal, WA82 Collaboration, Contribution to ICHEP
92 (Dallas, Texas, August 1992).}

\REF\vog{R. Vogt, S. J. Brodsky and P. Hoyer, SLAC-PUB-5827 (to be publ. in
Nucl. Phys. B).}
\REF\bmu{S. J. Brodsky and A. H. Mueller, Phys.~Lett. {\bf 206B} (1988) 285;
R. Vogt and S. Gavin, Nucl. Phys. {\bf B345} (1990) 104.}

There are several experiments on hadroproduction of
charm\refmark{\bia,\cha,\cot,\shi,\agu} which give
much larger cross sections at high $x$ than expected from leading twist QCD,
\ie, from Eq. \fact\ with a fragmentation function of the form \pete.
Recently, data with good statistics on $\pi^-A\to D+X$ \refmark{\app,\wa}
clearly shows that Eq. \fact\ underestimates the charm cross section already
at medium values of $x \gsim 0.2$. As shown in Fig. 6a,
the experimental $D$-meson distribution actually is similar to that
predicted by Eq. \fact\ for the charm {\it quark}\refmark{\app,\vog}. Thus
the fragmentation function $D(c\to D+X)$ must be assumed to be close to
$\delta(1-z)$ in order to get agreement between experiment and theory in
charm hadroproduction at medium $x$. Hadroproduced charm quarks appear
to fragment in a strikingly different way from what they do in $\epem$
annihilations, where the fragmentation function is to a good approximation
given by Eq. \pete. This implies a breakdown of the leading twist
factorization formula \fact.

The different form of the fragmentation function in $\epem$ induced, compared
to hadron induced, charm production has a natural
explanation\refmark{\vog,\bmu}.  In hadroproduction, the charm quark can
coalesce with light spectator quarks from the projectile, which move in the
same direction and with similar velocity as the charm quark. The charmed
hadron formed this way will have the same velocity as the charm quark,
resulting in a fragmentation function close to the $\delta$-function
suggested by the data.

\REF\barl{M. Aguilar-Benitez {\it et al.}, Phys. Lett. {\bf 161B} (1985)
400 and Z. Phys. {\bf C31} (1986) 491; S. Barlag {\it et al.}, Z. Phys. {\bf
C49} (1991) 555.}

The existence of coalescence between charm quarks and light valence quarks
from the projectile is suggested also by the ``leading particle
effect''. Those charmed hadrons that have a valence quark in common with
the projectile are experimentally found\refmark{\app,\wa,\barl} to have a
harder $x$-distribution. In Fig. 6b,c we show the size of this effect as
measured by E769\refmark\app, and as expected in a model with
coalescence\refmark\vog. From this model one can also see that the E769 data
is not very sensitive to the estimated contribution of intrinsic charm,
which is important only at the highest values of $x$.

\REF\bagl{C. Baglin, \etal, Phys.~Lett. {\bf 220B} (1989) 471.}
\REF\leit{D. M. Alde, \etal, Phys.~Rev.~Lett. {\bf 66} (1991) 2285; M. J.
Leitch, \etal, Nucl.~Phys.~{\bf A544} (1992) 197c.}
\REF\masa{T. Matsui and H. Satz, Phys.~Lett.~{\bf B178} (1986) 416.}

\REF\bad{J. Badier {\it et al.}, Z. Phys. {\bf C20} (1983) 101.}
\REF\kat{S.  Katsanevas {\it et al.},  Phys. Rev.  Lett. {\bf 60} (1988)
2121.}
\REF\ald{D. M. Alde {\it et al.}, Phys. Rev. Lett. {\bf 66} (1991) 133.}

\REF\hvs{P.  Hoyer, M. V\"anttinen, and U.  Sukhatme, Phys.  Lett. {\bf 246B}
(1990) 217.}

\chapter{The $A$ Dependence of Quarkonium Production}

The coalescence of heavy quarks with light comovers has an indirect effect
on the production of quarkonium states, such as the $\Jpsi$ and $\Upsilon$.
When many comoving light quarks are present, as in the fragmentation region
of heavy nuclei, the heavy quarks may preferentially coalesce with comovers
rather than bind to each other\refmark\bmu. This leads to a suppression of
quarkonium production in nuclear fragmentation regions, which has
been observed for the $\Jpsi$ in central heavy ion
collisions\refmark{\bagl}. A similar suppression is also observed in
``backward'' $(x<0)$  production of both the $\Jpsi$ and the $\Upsilon$ in
$pA$ collisions\refmark\leit, for which the alternative
explanation\refmark\masa\ in terms of a quark gluon plasma seems unlikely.

The forward $(x>0)$ production of the $\Jpsi$
has been measured with high statistics for both pion and proton beams on a
variety of nuclear targets.\refmark{\bad,\kat,\ald} In this case the effect
of coalescence should be small, since the $c\bar c$ state is produced in the
fragmentation region of a hadron, and has relatively few comovers. This data
nevertheless gives direct evidence\refmark\hvs\ for the breakdown of the
leading twist approximation at large $x$. In the factorized formula \fact,
the nuclear target $A$-dependence can only appear through the target
structure function $F_{b/A}(x_2)$, where $x_2$ is the momentum fraction
carried by the target parton $b$. Ratios of $\Jpsi$ cross sections at
different production energies but at the same $x_2$ should therefore be
independent of the nuclear number $A$. As shown in Fig. 7 for the NA3
data\refmark\bad\ on $\pi^-A\to \Jpsi+X$ at 150 and 280 GeV, the ratio of
cross sections on Hydrogen and Platinum does not in fact scale as a function
of $x_2$. Thus the leading twist factorization \fact\ fails. A similar
result was obtained by combining $pA$ data from NA3 and E772\refmark\ald.

The failure of factorization occurs at the smallest
values of $x_2$, \ie, for large Feynman $x$ of the $\Jpsi$, since $x_2\simeq
M^2_\Jpsi/xs$. In this region of $x$ the nuclear target dependence of the
$\Jpsi$ cross section also is not linear in $A$, as expected for hard QCD
processes which occur incoherently off all partons in the nucleus. If the
$\Jpsi$ cross section is parametrized as
$$\sigma_{hA} = \sigma_{hN} A^\alpha\eqn\adep$$
one finds\refmark{\bad,\kat,\ald} that $\alpha=0.7\ldots 0.8.$

\REF\shad{For discussions of nuclear structure function shadowing in QCD,
see, e.g., L. L. Frankfurt and M. I. Strikman,
Nucl. Phys. {\bf B316} (1989) 340;
A. H. Mueller and J.-W. Qiu
Nucl. Phys. {\bf B268} (1986) 427;
S. J. Brodsky and H. J. Lu, Phys. Rev. Lett. {\bf 64} (1990) 1342;
V. Barone, M. Genovese, N. N. Nikolaev, E.Predazzi and B. G. Zakharov,
University of Torino preprint DFTT 14/92.}

\REF\arn{M. Arneodo, CERN-PPE/92-113 (June 1992, submitted to Phys.
Reports.)}

At small values of $x_2$, one does expect $\alpha\lsim 1$ due to parton
shadowing\refmark\shad, as observed\refmark\arn\ in deep inelastic lepton
scattering (DIS) and in lepton pair production (DY). However, shadowing
appears to be a $Q^2$-independent, leading twist effect associated with the
nuclear structure function. It thus cannot account for the breakdown of
factorization observed in $\Jpsi$ production. Moreover, the shadowing
effect seen in DIS and in DY is a fairly small, $10\ldots 30$\% effect,
whereas the suppression of $\Jpsi$ production amounts to a factor $\sim 3$
for large nuclei.

\REF\gm{S.  Gavin and J.  Milana, Phys.  Rev.  Lett. {\bf 68} (1992) 1834.}

\REF\qu{E.  Quack, Heidelberg preprint HD-TVP-92-2 (June 1992).}

\REF\ff{S. Frankel and  W. Frati, University of Pennsylvania preprint
UPR-0499T (May 1992).}

\REF\emc{EMC Collaboration, A.  Arvidson \etal, Nucl.  Phys.
{\bf B246} (1984) 381;
R.  Windmolders, Proc. of the 24.  Int.  Conf. on High
Energy Physics, (Munich 1988, Eds.  R.  Kotthaus and J.  H.  K\"uhn, Springer
1989), p. 267.}

\REF\dy{D.  M.  Alde \etal, Phys.  Rev.  Lett. {\bf 64} (1990) 2479.}

\REF\bh{S. J. Brodsky and P. Hoyer, SLAC-PUB-5935 (September 1992,
submitted to Phys. Lett. B).}

It was recently suggested \refmark{\gm,\qu,\ff} that the nuclear
suppression of $\Jpsi$ production could be due to energy loss of the
incoming and outgoing partons while propagating through the nucleus.
However, very little energy loss is observed for the struck quark in the DIS
process\refmark\emc, as well as for the incoming quark in the DY
reaction\refmark\dy. This shows that high energy quarks suffer only
insignificant energy loss in the nucleus. Actually, this fact is a direct
consequence of the uncertainty principle, which forbids enhanced energy loss
from multiple collisions occurring within the formation zone of the radiated
gluons\refmark\bh. Repeated radiation from collisions separated by a time
interval $\Delta t$ is allowed only provided $\Delta E \Delta t\gsim 1$, \ie,
when the energy difference resulting from the gluon radiation $\Delta E$ is
big enough for the multiple scatterings to be resolved. In a nucleus,
$\Delta t\lsim R_A$, where $R_A$ is the nuclear radius. The emission of a
gluon with transverse momentum $p_\perp$ and energy $E_g$ results in
$\Delta E\simeq p_\perp^2/2E_g$. Hence normal, soft collisions in the
nucleus with $\VEV{p_\perp^2}\sim 0.1$ GeV$^2$ can only lead to a {\it
finite} energy loss in the laboratory frame,
$$E_g\lsim\half \VEV{p_\perp^2} R_A \lsim 1.5~\GeV.\eqn\eloss$$
For high energy partons this fixed energy loss
is not significant, and it cannot explain the nuclear suppression of $\Jpsi$
production.

When the fixed energy loss \eloss\ is small compared to the energy of the
$\Jpsi$, we expect to see Feynman scaling (in $x$) of the $\Jpsi$
cross section. This is indeed observed in the data (\cf\ Fig.
7). Hence it is most natural to discuss the $\Jpsi$ $x$-distribution from the
point of view of the projectile wave function. This brings us back to our
earlier discussion of heavy quark production at large $x$. The
$A$-dependence of the $\Jpsi$ data is indeed one of the strongest arguments
for the relevance of the new QCD limit \mlimit, where $\M^2(1-x)$ is held
fixed as $\M^2\to \infty$ and $x\to 1$.

\REF\vbh{R. Vogt, S. J. Brodsky, and P. Hoyer,
Nucl. Phys. {\bf B360} (1991) 67.}

In Section 4 we saw that a virtual heavy quark pair with $x$ near 1 can be
put on its mass shell (produced) by a soft gluon scattering off the stopped
light quarks. If $1-x=\mu^2/\M^2$, the light quark distribution has a
transverse size $1/\mu$, and the hardness of the target interaction is
$Q^2=\mu^2$ (\cf\ Eq. \pperp). Hence for small $\mu\sim \Lambda_{QCD}$, the
scattering will be surface-dominated in a big nucleus, and the cross
section gets an $A^{2/3}$ nuclear target dependence. There should thus be a
smooth transition from an $A^1$ behavior at small $x$ (hence large $\mu$) to
an $A^{2/3}$ behavior at large $x$, in agreement\refmark\vbh\ with the data.

The NA3 data\refmark\bad\ was analyzed assuming the existence of two
components in the cross section, a ``hard'' component with $\alpha=.972$ in
\adep\ and a ``diffractive'' component with $\alpha=.77$ (for the $\pi^-A$
data) or $\alpha=.71$ (for the $pA$ data). A quantitative fit showed that the
hard component was in good agreement with expectations from the $GG\to
c\bar c$ and $q\bar q\to c\bar c$ fusion processes. Hence the
``diffractive'' component, which dominates for $x\gsim 0.6$, appears to be an
excess over the leading twist QCD fusion contribution. Note that
explanations of the anomalous nuclear dependence in terms of energy loss or
breakup of the $\Jpsi$ would not lead to any excess in the large $x$
production cross section on light targets.

Because the target interaction in the $x\to 1$ limit \mlimit\ can be soft
($\ell_\perp\sim \mu$ in Fig. 4), the average transverse momentum $\VEV{\kp}$
of the $\Jpsi$ should decrease from $\VEV{\kp}=\order{m_c}$ at low $x$ to
$\VEV{\kp}=\order{\mu}$ at large $x$. This effect has been seen in the
$\Jpsi$ data\refmark\bii. Furthermore, the anomalous $A$-dependence, \ie,
$\alpha<1$ in Eq. \adep, is observed\refmark{\bad,\kat} only at low $\kp$. A
similar decrease of $\VEV{\kp}$ with $x$ is expected also in open charm
production. In addition, coalescence between heavy and light quarks occurs
mainly for heavy quarks with low $\kp$, comparable to the light quark
transverse momenta. This also tends to decrease the $\VEV{\kp}$ of
charmed hadrons at large $x$.

For $\Upsilon$ production, the available data is not at large enough $x$ to
make $\mu^2=(1-x)\M^2$ small. Hence the corrections to the leading twist
results should not be very significant. The $A$-dependence of the $\Upsilon$
cross section is in fact found\refmark\leit\ to be much closer to $A^1$ than
is the case for the $\Jpsi$.

\chapter{Concluding Remarks}

We have discussed the corrections to the twist expansion of PQCD for hard
processes in the kinematic region where some of the constituents (or
hadrons) carry a large fraction $x \to 1$ of the available energy. We
found that the limit \xlimit, where the hard scale $Q^2$ (or $\M^2$) is
comparable to the scale $\Lambda_{QCD}^2/(1-x)$, is particularly
interesting. The production mechanism of a heavy quark pair of mass $\M$
becomes peripheral in this limit, with the hardness $\mu^2$ of the target
interaction being given by
$\mu^2=\M^2(1-x)$. For $\mu \lsim \Lambda_{QCD}$ this means that the
interaction in the target is soft. In this situation the factorization
formula \fact, according to which the scattering occurs incoherently off
single partons in the target, breaks down. In particular, it implies
scattering from the surface of large nuclei, and a consequent $A^{2/3}$
nuclear target dependence.

On the other hand, only very compact Fock states of the projectile, where
the transverse separation of the light valence quarks is of order $1/\M$,
can participate in the production. For large $x$ {\it several}
valence partons of the projectile are involved in the heavy quark production
process, again breaking the leading twist, single parton factorization
\fact. In this respect the dynamics is similar to that in the exclusive
($x=1$) limit. The hard production vertex can be calculated in PQCD, given
the longitudinal momentum distributions of the quarks in the transversally
compact Fock states of the projectile. So far, this calculation has only
been carried out\refmark\bhmt\ in a scalar QED model, however.

Phenomenologically, significant deviations from the leading twist
approximation have been seen at large $x$ in deep inelastic lepton
scattering\refmark\vm, in lepton pair production\refmark{\naten,\cp}, in
$\Jpsi$ hadroproduction\refmark{\bad,\kat,\ald} and in open charm
hadroproduction\refmark{\bia,\cha,\cot,\shi,\agu}. Recently, firm evidence
was obtained\refmark{\app,\wa} that there are significant corrections to the
leading twist PQCD cross section for open charm hadroproduction even at
moderate $x \gsim 0.2$. The most natural explanation for this effect is the
coalescence of the charm quark with co-moving light spectator
quarks\refmark{\vog,\bmu}. Coalescence is a soft process, which is not
directly calculable in PQCD. It allows the charm quark to maintain its
velocity during hadronization, and can thus effectively be described by a
fragmentation function $D(z)\simeq\delta(1-z)$.

\REF\bein{H. U. Bengtsson and G. Ingelman, Comp. Phys. Comm. {\bf 34}
(1985) 231.}
\REF\ben{H.-U. Bengtsson and T. Sj\"ostrand, Comp. Phys. Comm.
{\bf 46} (1987) 43.}

A significant change in the velocity of a heavy quark always requires large
momentum transfer -- hence this is a process that is associated with the
hard production vertex, not with soft hadronization. Some
fragmentation schemes based on the string model\refmark{\bein,\ben} give rise
to charmed hadron distributions which are considerably harder than the
distribution of the charm quark\refmark{\agu,\wa}. Such mechanisms seem to go
beyond the realm of soft physics for which the models were intended, and are
thus unreliable. The methods that we have described above illustrate how the
cross section of heavy quark production at the largest values of $x$ can be
calculated in QCD. While the size of the new, intrinsic contributions to
charm production has not yet been determined from the theory,
phenomenological estimates\refmark{\vog,\vbh}
show that they are important for $x \gsim 0.5$. Several
experiments have in fact reported larger charm hadroproduction cross sections
at high $x$ than would be expected from the leading twist approximation
\fact. These data await confirmation by upcoming, high statistics
experiments.

\vskip 1cm
{\bf Acknowledgements}

The material presented here is the result of
collaborations and discussions with, in particular, Stan Brodsky, Vittorio
Del Duca, Al Mueller, Wai-Keung Tang and Ramona Vogt. I am also grateful to
the organizers of the Zakopane School for their invitation and warm
hospitality.

\refout
\vskip 1cm
\centerline{FIGURE CAPTIONS}
\medskip
\item{\rm Fig. 1} The $x\to 1$ limit of a hadron structure function is
generated by perturbative gluon exchange. The transverse size $r_\perp$ of
the initial Fock state is small, $r_\perp^2\sim (1-x)/p_\perp^2$. The
virtual state with large $x$ has a larger size $R_\perp\sim 1/p_\perp$, due
to the transverse motion of the stopped quark carrying the small momentum
fraction $1-x$.

\item{\rm Fig. 2} Deep inelastic lepton-hadron scattering. (a) A leading
twist diagram. (b) A higher twist diagram.

\item{\rm Fig. 3} Leading twist diagrams in heavy quark production at fixed
momentum fraction $x$ of the heavy pair. (a) A lowest order $GG\to c{\bar
c}$ diagram. (b) A higher order $\bar q G\to\bar q Q\bar Q$ diagram. The
gluons have virtualities ranging up to the mass scale $\M$ of the heavy
pair. There is no restriction on the transverse size of the initial Fock
state.

\item{\rm Fig. 4} Leading order diagrams in the limit \mlimit. In the
extrinsic diagram (a) the produced heavy quark pair couples directly to only
one parton in the projectile. Diagram (b) is intrinsic, as
the  $Q{\bar Q}$ pair couples to two partons.  In
the standard limit \qlimit, both diagrams (a) and (b) would be classified as
higher twist, and would be suppressed by $1/{\cal M}^2$. The gluon from the
target has a virtuality of $\order{\mu}$, while the two other gluons have
virtualities of $\order{\M}$. The transverse sizes $r_\perp$ and $R_\perp$
of the initial and final Fock states are as in Fig. 1, and the size of the
$Q\bar Q$ pair is $h_\perp\sim 1/\M$.

\item{\rm Fig. 5} In an intrinsic process the intermediate state indicated
by the vertical dashed line can be on-shell, resulting in an imaginary part
of the amplitude at leading order.

\item{\rm Fig. 6} $x_F$ distributions\refmark\vog\ of $D$ mesons produced in
$\pi^-A$ collisions at 250 GeV/c (E769 data\refmark\app). (a) The
solid line shows the prediction of leading twist QCD, using a
Peterson\refmark\pet\  fragmentation scheme for $c\to D+X$ which fits the
data on $D$ production in $e^+e^-$ annihilations. The dashed curve shows the
effect of adding an estimated intrinsic charm contribution to the
fusion cross section. The dot-dashed curve results from
fusion + intrinsic charm at the quark level, {\em i.e.}, for
$\delta$-function fragmentation. (b) $D^-$ production, for which
the curves include both the fusion and intrinsic production mechanisms. The
solid curve is obtained with Peterson fragmentation and the dashed curve
with $\delta$-function fragmentation. (c) As in (b), for
$D^+$ production.

\item{\rm Fig. 7} The ratio $R=A\sigma(pp\to\Jpsi+X)/ \sigma(pA\to\Jpsi+X)$
of inclusive $\Jpsi$ production cross sections on Hydrogen and
Platinum\refmark\bad. In (a) the ratio is plotted as a function of the
Feynman $x_F$ of the $\Jpsi$, and in (b) as a function of the momentum
fraction $x_2$ of the target parton\refmark\hvs.

\bye